\documentclass[runningheads]{llncs}
\usepackage[T1]{fontenc}
\usepackage{graphicx}
\usepackage{hyperref}
\usepackage{color}

\usepackage{amsmath}  %
\usepackage[english]{babel}

\usepackage{listings}

\definecolor{codegray}{rgb}{0.4,0.4,0.4}
\definecolor{backcolour}{rgb}{0.98,0.98,0.98}
\definecolor{codepurple}{RGB}{127,0,85}
\definecolor{darkolivegreen}{rgb}{0.33,0.42,0.18}

\begin{document}

\title{Iconographic Classification and Content-Based Recommendation for Digitized Artworks}

\titlerunning{Iconographic Classification and Content-Based Recommendation for \ldots}

\author{Krzysztof Kutt\orcidID{0000-0001-5453-9763} \and
Maciej Baczy\'{n}ski
}

\authorrunning{K. Kutt and M. Baczy\'{n}ski}

\institute{Department of Human-Centered Artificial Intelligence, Institute of Applied Computer Science, Faculty of Physics, Astronomy and Applied Computer Science, Jagiellonian University, prof. Stanis\l{}awa \L{}ojasiewicza 11, 30-348 Krak\'{o}w, Poland\\
\email{krzysztof.kutt@uj.edu.pl}}

\maketitle              %

\begin{abstract}
We present a proof-of-concept system that automates iconographic classification and content-based recommendation of digitized artworks using the Iconclass vocabulary and selected artificial intelligence methods.
The prototype implements a four-stage workflow for classification and recommendation, which integrates YOLOv8 object detection with algorithmic mappings to Iconclass codes, rule-based inference for abstract meanings, and three complementary recommenders (hierarchical proximity, IDF-weighted overlap, and Jaccard similarity).
Although more engineering is still needed, the evaluation demonstrates the potential of this solution:
Iconclass-aware computer vision and recommendation methods can accelerate cataloging and enhance navigation in large heritage repositories.
The key insight is to let computer vision propose \emph{visible} elements and to use symbolic structures (Iconclass hierarchy) to reach \emph{meaning}.

\end{abstract}

\section{Introduction}
\label{sec:intro}

Cultural heritage (CH) encompasses resources inherited from past generations that communities recognize as expressing evolving values, beliefs, and knowledge.
Digitization has transformed access to these CH resources, including artworks, making them available on a large scale, but at the cost of losing the interpretive context provided by experts in museums and galleries.
While descriptive metadata (date, place, author) support basic retrieval, iconographic access depends on identifying \emph{what} is depicted and \emph{how} it is symbolized---a task traditionally performed by experts.
Computational systems must therefore help reconstruct the meaning.

Recent surveys (e.g.,~\cite{caron2023cvch}) emphasize that computer vision (CV) is increasingly embedded in CH pipelines---from acquisition and documentation to interpretation and dissemination---under constraints unique to historical materials (rarity, aging, small datasets).
This ``digital heritage'' agenda advocates cross-disciplinary design of models and interfaces that respect conservation realities while enabling large-scale analysis.

However, as highlighted by~\cite{wu2023metadata}, machine learning is only as good as its training data, and the CH domain suffers from sparse and heterogeneous labels.
The authors advocate for transparent rules, shared vocabularies, and realistic expectations about what can (and cannot) be automated
with a \emph{scale} and \emph{semantics} bottleneck in mind:
millions of images require consistent, interpretable descriptions, and thematic navigation.
This cautions against free-form labels that derail cross-collection interoperability and reinforces the strategic role of Iconclass classification system and other standardized vocabularies~\cite{brandhorst2024babel}.

This paper reports a prototypical CARIS system (``Classification and recommendation for the Iconclass system'') that
(a) proposes Iconclass codes based on visual content analysis and
(b) recommends thematically related items.
The work responds to the growing emphasis on machine-learning support for heritage interpretation (cf., e.g., ``AI and Artworks: Object Detection, Image Classification and Iconographic Analysis'' workshop\footnote{\url{https://indico.global/event/15085/} (accessed: 12.02.2026)}).
The goal is to accelerate expert workflows and enrich user discovery rather than to replace curatorial judgment.

The remainder of the paper is structured as follows.
Related works are summarized in Sect.~\ref{sec:sota}.
The proposed system is described in Sect.~\ref{sec:overview} and evaluated in Sect.~\ref{sec:evaluation}.
Sect.~\ref{sec:summary} concludes the paper.

\section{Related works}
\label{sec:sota}

\subsection{Iconography and Iconclass}
\label{sec:sota:iconclass}

Iconography studies motifs, attributes, narratives, and their symbolic meanings.
Iconclass~\cite{couprie1983iconclass} is the most widely used controlled vocabulary for such analysis in (mainly Western) art:
a hierarchical alphanumeric code system with optional qualifiers for nuanced meaning.
The top level comprises ten groups (0-9: from abstract art and nature to religion, history, literature, and classical mythology).
Each additional character refines meaning: digits alternate with capital letters, %
and deeper levels encode increasing specificity.
Beyond the tree, Iconclass supports
bracketed text (for species names, proper names, numbers),
keys (contextual qualifiers such as ``heraldry''),
doubling of letters (e.g., gendered or indoor/outdoor contrasts in some branches),
and structural digits (position-dependent semantics used to align life events across different persons).
Together, these mechanisms allow for coarse and highly granular descriptions of depicted scenes and symbols~\cite{brandhorst2024babel} (some examples are provided in Sect.~\ref{sec:overview}-\ref{sec:evaluation}).
The presented structure induces a semantic distance:
codes sharing long prefixes are thematically close even if not identical, which will be crucial for recommendation task.

Nowadays, the system is digitized and available via the website \url{https://iconclass.org/} and a dedicated Python package\footnote{\url{https://pypi.org/project/iconclass/}}.
The authors provide not only a description of the taxonomy, but also additional digital resources related to the system, such as the Iconclass AI Test Set (${\approx}87$k images with Iconclass codes available at \url{https://iconclass.org/testset/}). 
The main limitation of this data set is the small size of the images, which was chosen to save space---the maximum size of 500 pixels on the longer side is too small to perform effective visual recognition.
However, it can still be used to evaluate the recommendation module of the CARIS prototype described in this paper.

\subsection{Computer vision methods in CH pipelines}
\label{sec:sota:cv}

Computer vision has become a foundational technology in the analysis, preservation, and documentation of CH materials.
Its ability to detect, classify, and localize visual elements enables scalable interpretation of artworks and other artifacts.
This shift aligns with a wider trend of integrating deep learning into heritage pipelines, where automated vision systems assist experts by accelerating tasks such as damage detection, motif extraction, and object identification~\cite{guo2025ai}.

Within this landscape, YOLO (You Only Look Once) has emerged as one of the most influential single-stage object detection architectures due to its speed, accuracy, and adaptability~\cite{ali2024yolo}.
Recent studies demonstrate that YOLO is increasingly being applied in cultural heritage applications, especially where fine-grained visual cues are critical~\cite{chen2025damage,guo2025ai,wang2025mural}.
YOLO models have also been used in iconography, but these have been very limited attempts, e.g.,~\cite{monleon2025msc,rizvi2025msc}.

\subsection{Automatic iconographic classification}
\label{sec:sota:classification}

Early deep-learning efforts targeted focused iconographies, e.g., saints in Christian paintings.
One of the first attempts introduced a dataset and CNN that achieved about 70\% F1 despite subtle inter-class differences, demonstrating feasibility but highlighting the challenge of fine-grained, visually similar classes typical of art history~\cite{milani2020iconography}.

Scaling to the full Iconclass hierarchy, hierarchical multi-label classification over ${>}20$k Iconclass concepts and ${\approx}478$k images was proposed~\cite{springstein2024visualnarratives}, leveraging language models to synthesize textual descriptions from Iconclass keywords and pretraining a vision-language model.
Their transformer explicitly encodes taxonomy constraints and significantly outperforms flat baselines---evidence that hierarchy-aware architectures are key for iconography at scale.

Iconclass classification was also attempted as cross-modal retrieval using artworks' titles (multi-lingual) and images~\cite{banar2023iconclass}.
Textual signals often dominate, but visual features contribute complementary evidence and enable cross-lingual transfer (Dutch-English) without quality loss.

\subsection{Artwork recommendation}
\label{sec:sota:recommendation}

Recommendation systems are applications that are used to analyze user preferences and propose the most relevant data.
Such systems can be found in various every-day scenarios, e.g., social networks suggest content to view, e-commerce sites recommend products that users are likely to enjoy~\cite{raza2026rs}.

Early recommendation systems in the CH domain relied heavily on metadata---such as artist, medium, style, or period---but these approaches provided only limited semantic coverage.
A key advance came from integrating visual features extracted by CV methods, e.g., it was demonstrated that combining curated metadata with DNN-derived latent visual features substantially improves recommendation performance for physical paintings on commercial platforms such as UGallery\footnote{\url{https://www.ugallery.com/}}~\cite{messina2019artrec}.

The main limitation of visual-based recommendations was indicated in the introduction (Sect.~\ref{sec:intro}): they are not based on structured vocabularies, which offer unique advantages, as they encode thematic and symbolic meaning, enabling recommendation engines to group artworks not just by visual clues but by iconographic content.
To the best of authors' knowledge, there are no reported works explicitly combining Iconclass with recommender systems.
However, existing work on hierarchical classification~\cite{springstein2024visualnarratives} and semantic-aware recommenders~\cite{huang2019taxonomy,zhang2019taxonomy} suggests strong compatibility between semantic taxonomies like Iconclass and content-based recommendation.

\subsection{IDF and Jaccard similarity}
\label{sec:sota:similarity}

During the work on the CARIS system, it became necessary to introduce a recommendation function that would take into account the rarity of codes within the set under consideration. The solution was to use the IDF (inverse document frequency) metric, which assigns weights to individual codes depending on how many documents in the set they appear in, while rewarding the rarer ones:
\[
\text{IDF}(c) = \log \left( \frac{N}{n_c} \right)
\]
where, $N$ denotes the number of all documents, and $n_c$ denotes the number of documents containing the considered code.

IDF usually goes paired with TF (term frequency) metric, creating TF-IDF, which allows for identifying terms relevant within a document against the background of their entire collection~\cite{geeksforgeeks_tfidf}.
Such a procedure facilitates the identification of the subject matter of a given document, as well as finding similar ones in the collection.
However, the use of TF in the reported work is not possible, as each code within a given set can only occur once, so it is not possible to calculate their frequency of occurrence.

In the CARIS system, it was also necessary to take into account the ratio of repeating codes between the considered images to the sum of all codes assigned to them.
This is made possible by the Jaccard similarity coefficient~\cite{geeksforgeeks_jaccard}, which defines the similarity between sets A and B as:
\[
J(A, B) = \frac{|A \cap B|}{|A \cup B|}
\]

\section{System overview}
\label{sec:overview}

We propose a prototypical system called ``CARIS: Classification and recommendation for the Iconclass system''.
Given a digitized artwork, it implements a four-stage pipeline~\cite{baczynski2025msc}:
\begin{enumerate}
    \item detects visible objects with YOLO,
    \item proposes Iconclass codes consistent with those detections thanks to mappings of Iconclass codes to YOLO tags,
    \item infers abstract codes where possible,
    \item and recommends thematically related artworks using only Iconclass codes (purely content-based recommendation, without user history).
\end{enumerate}

The system is delivered as a Python package with dedicated modules for I/O, classification (steps 1–3 from the pipeline), and recommendation (step 4).
The orchestration wrapper (\emph{classify\_and\_recommend()} function)  chains detection, code mapping and inference, recommendation, and returns (a) proposed Iconclass codes and (b) three recommended images (one per method, as described in Sect.~\ref{sec:overview:recommendation}).
Design follows SRP (single responsibility principle) and DRY (don't repeat yourself), avoids global state, and uses the official Iconclass Python resources for traversals, keywords, and code metadata.

The entire CARIS code is provided in a dedicated repository at \url{https://gitlab.geist.re/pro/caris-iconclass}.

\subsection{Classification}
\label{sec:overview:classification}

We employ a YOLOv8 to extract unique object \emph{labels} from an artwork. Duplicates---e.g., if two horses are visible in the image---are removed as Iconclass assumes that a single code is assigned to a given type of object, regardless of the number of its occurrences.
Then, it is necessary to map the detected \emph{labels} to the Iconclass system \emph{codes}.
We implement two complementary mapping strategies:
keyword-based set matching (primary)
and description search (secondary).

The first method is based on the \emph{.keywords()} method implemented in the Iconclass library, which returns a list of \emph{keywords} associated with the \emph{code}.
These are usually simple English terms, such as ``dog'' and ``dye''.
Such a combination may seem unusual, but there is a code ``94L53'' that links them together: ``Hercules discovers Tiryns' famous dye: the muzzle of Hercules' dog is stained with purple after it has bitten into a mollusc''.
Such a highly detailed granularity of the Iconclass system requires appropriate search methods to retrieve very specific codes without overlooking more general ones.
We propose a three-pass mapping algorithm with search relaxation:
\begin{enumerate}
    \item \emph{Exact set match}.
    The algorithm begins by searching for the narrowest, and thus the best and most accurate match, i.e., it searches for an exact match between the set of \emph{labels} from YOLO and the set of \emph{keywords} for the Iconclass code. If such matches are found, the algorithm terminates and returns the Iconclass \emph{codes} found.
    
    \item \emph{Labels $\subseteq$ keywords}.
    If no \emph{code} is found in the first step, then the relaxation of requirements follows.
    In the next step, equality between the \emph{labels} and \emph{keywords} sets is no longer required.
    It is sufficient for the first set to be a subset of the second.
    The motivation for this step is the observation that individual Iconclass \emph{codes} may have many different \emph{keywords}, and some of the objects (\emph{labels}) corresponding to them may not have been correctly identified by the YOLO model or may not have been directly represented in the analyzed image.
    However, such relaxation may cause problems with images that contain simple scenes in which few elements are recognized.
    For example, if only a dog is recognized in the analyzed image, the user will receive all Iconclass \emph{codes} containing the word ``dog'' among their \emph{keywords}.
    These will include \emph{codes} related to history, the Bible, literature, or mythology, because dogs were involved in the events to which they refer.
    Therefore, it is important to start the search with the more restrictive criterion described in step 1.
    
    \item \emph{Singleton searches per label}.
    Regardless of whether any \emph{codes} were found, the algorithm may perform another (optional) step, in which the search is conducted based on each of the identified \emph{labels} separately.
    This search significantly increases the total number of \emph{codes} returned as a result, because the chance of finding a single \emph{label} among \emph{keywords} associated with \emph{codes} is very high.
    It also allows individual depicted objects to be assigned corresponding \emph{codes}, which is common practice in classification using the Iconclass system.
\end{enumerate}

The second evaluated mapping method was based on linking the name of the detected object (\emph{label}) with \emph{descriptions} of individual codes instead of comparing it with \emph{keywords}.
Even in the initial testing phase, this method proved to be less accurate, which is why it will not be described in detail here.
The implementation of this approach will be analogous to the keyword-based method mentioned above, with the difference that instead of the \emph{.keywords()} method, \emph{.texts()} will be used.

The mechanisms presented above are only capable of suggesting codes that refer to objects literally depicted in the image. Abstract terms such as justice cannot be clearly identified with a single object that could be detected by the YOLO model. At least a partial solution to this problem is provided by the fact that many of these symbols are based on the common occurrence of several objects that can be directly detected. Justice, for example, can be depicted as a woman with blindfolded eyes, holding scales and a sword. If all these elements are present in the image, there is a significant probability that its meaning is related to justice. Similar logic can be applied to specific events or group scenes. If a deer, a dog, a horse, and a human can be seen, the artwork potentially depicts hunting.
In order to improve the quality of classification and to include abstract concepts, the system implements the above-described logic of inferring new codes based on other already identified codes.
This lightweight rule engine is based on a transparent JSON for auditability. The rules can be bootstrapped with curator input or assisted by generative models (used only for rule suggestion).

Iconclass's uneven coverage (e.g., rich religious subtrees; many dog-related narrative codes) can yield many plausible codes from a single common object.
For example, if only a dog is detected in an image, the method described above in step 1 (exact set match) will return a total of 6 codes (see Listing~\ref{lst:dog}).
This leads to the need for post-filtering, which will limit the number of codes.

Two deterministic reducers are available:
(1) intersection of codes sets returned by keyword- and description-based mappings,
and (2) a shortest title heuristic that prefers the most specific code whose label minimally extends the detected object term (often the generic object code).
In practice, both methods demonstrated low accuracy, and therefore are not automatically launched in the main function of the system.

This task can be performed more effectively using generative models.
However, it is important to use them only as a selector filtering redundant Iconclass codes, and not as a code generator, because in the latter situation they often hallucinate nonexistent codes, which is consistent with broader limitations documented in~\cite{wu2023metadata}.

\begin{lstlisting}[
  caption={Iconclass codes with exactly one keyword = ``dog''.},
  label=lst:dog,
  float=t,
]
11H(CRISPIN & CRISPINIAN)69 dogs and/or wild animals do not touch the bodies
34B11 dog
43A3746 dogs (circus performance)
43C2181 dogs (racing)
46E31 dog (as messenger)
73F215321 Peter sends a dog into the house to summon Simon to come out
\end{lstlisting}

\subsection{Recommendation}
\label{sec:overview:recommendation}

Given the Iconclass code set for a query image, we employ three recommenders that reflect complementary semantic principles:
\begin{enumerate}
    \item \emph{Hierarchy-based similarity}, consistent with hierarchical modeling~\cite{springstein2024visualnarratives}. Use score $1.0$ for identical codes, $0.5$ for codes sharing an immediate parent (or parent-child), and $0.25$ for sharing a grandparent (or two-step parent-child). Then sum over codes. This explicitly exploits Iconclass's tree semantics, mirroring the advantages seen in hierarchical classifiers.
    \item \emph{IDF-weighted overlap}, where rare codes carry greater semantic weight~\cite{milani2020iconography}.
    Compute IDF per code over the corpus, sum IDF for shared codes, and optionally exponentiate by \emph{idf\_impact} so that rare, diagnostic codes dominate over common object codes.
    \item \emph{Jaccard similarity}, robust against artworks with many codes. Use on code sets to counter biases toward images with many codes and to favor tight thematic overlaps.
\end{enumerate}

\section{Evaluation}
\label{sec:evaluation}

\subsection{Classification}
\label{sec:evaluation:classification}

\begin{figure}[t]
    \centering
    \includegraphics[width=.75\textwidth]{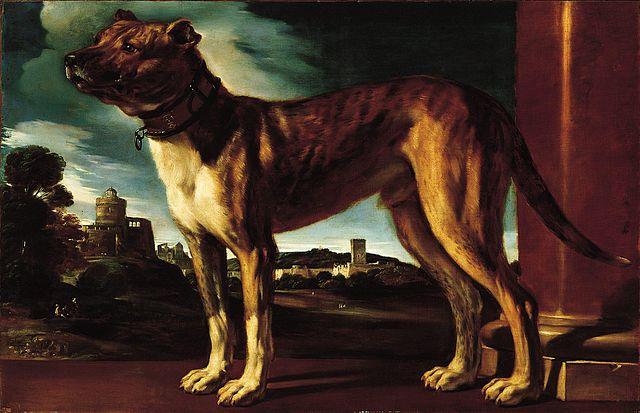}
    \caption{``The Aldrovandi Dog'' by Guercino. Wikimedia Commons, public domain.}
    \label{fig:dog}
\end{figure}

\begin{figure}[t]
    \centering
    \includegraphics[width=.75\textwidth]{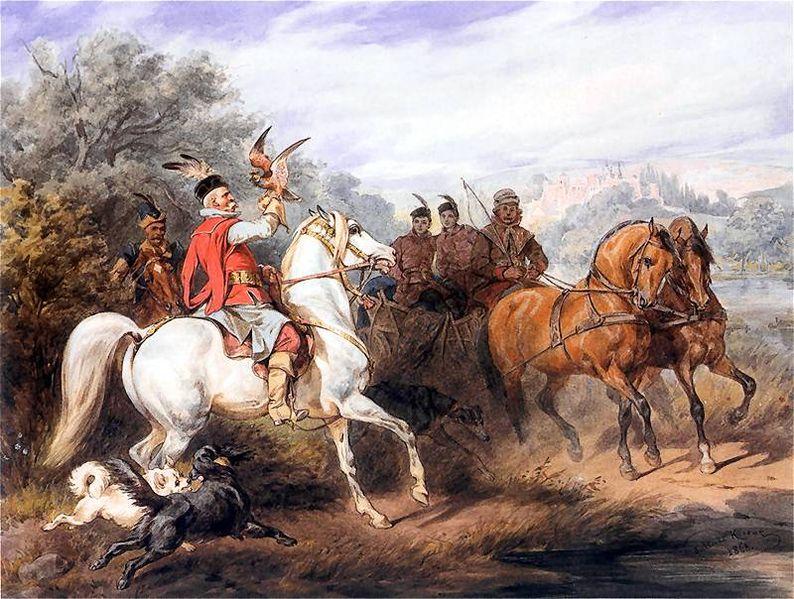}
    \caption{``The Hunting With Falcon'' (Polish: ``Wyjazd na polowanie z sokołem'') by Juliusz Kossak. Wikimedia Commons, public domain.}
    \label{fig:hunt}
\end{figure}

The classification module was evaluated with public domain images from Wikimedia Commons.
Two examples are shown to illustrate how it works.
The first is a single-object portrait of a dog (see Fig.~\ref{fig:dog}).
YOLO correctly detected the dog, which led to the retrieval of 6 Iconclass codes for which the only keyword was dog (see Listing~\ref{lst:dog}).
The reducer correctly selects one code: ``34B11 dog''.

The second example is a rich scene of falconry hunt (see Fig.~\ref{fig:hunt}), with horses, humans, dogs, and prey.
YOLO missed falcon and dogs, which degraded the label set to horse and human being.
With this starting point, the system correctly found the code ``46C13141(+78) horse (+ man and animal)'' connecting both keywords.
Switching on the optional step \emph{singleton searches per label} (see Sect.~\ref{sec:overview:classification}), which searches for codes for each identified object separately, led to an explosion of codes, as there are as many as 1231 codes with the keyword ``human being'' in the Iconclass system.
Many of these codes refer to depictions of various animals with humans, e.g., ``25FF24(MUSK-DEER)(+78) hoofed animals: musk-deer - FF - fabulous animals (+ man and animal)''.

The problem of code explosion is the challenge related to the design of the Iconclass system itself.
It results from the fact that some Iconclass codes have keywords assigned in a non-obvious way or have too few keywords assigned to them for effective classification based on them.
An example of such a code is ``43CC114(+423) hunter - CC - female hunter, huntress (+ hunting with horses)'', which can be simplified to ``a hunter hunting with a horse''.
However, its only keyword is ``horse''.
It has no assigned keywords related to hunting or hunters, which might seem intuitive.
As a result, whenever the system tries to find a code describing a horse, it also returns ``43CC114(+423)'', which may have no connection to the current image.
This problem is partially solved by the proposed reducers.
In particular, the generative model consistently removed false positives and preserved salient codes (e.g., reducing multiple dog-related codes to ``34B11 dog'' for a canine portrait).

Manual tests on historical paintings show that detection recall is the principal bottleneck.
If an element important to the image is omitted or misidentified, the assigned Iconclass codes will not reflect the content of the image.
This was particularly true for animals, with a dog mistaken for a bear and a cow mistaken for a horse, among others.
From the perspective of the classification system, such a mistake can change the meaning of the image from a noble hunt to a rural landscape.
Fine-tuning detectors on a curated label set aligned to Iconclass keywords is a priority for future work.

Despite the problems mentioned above, in most cases, the system fulfills its objectives and correctly returns codes corresponding to objects present in the image. Its operation is significantly improved by mechanisms for inferring new codes and reducing returned codes, enabling more accurate results. The proper functioning of the system confirms the validity of its initial assumptions.

\subsection{Recommendation}
\label{sec:evaluation:recommendation}

The recommendation module was evaluated with the Iconclass AI Test Set (see Sect.~\ref{sec:sota:iconclass}).
${\approx}87$k graphics allow for a reliable recommendation test, and the fact that Iconclass codes have been assigned by experts makes the evaluation independent of the effectiveness of the classifier.

\begin{figure}[t]
    \centering
    \includegraphics[width=.39\textwidth]{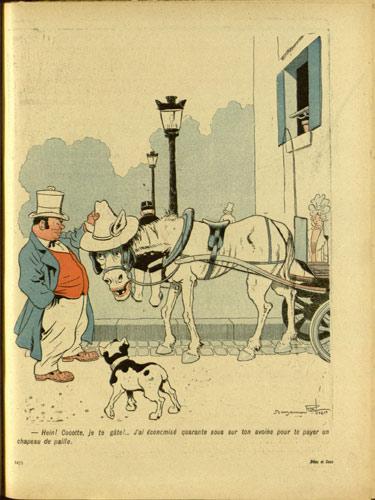}
    \includegraphics[width=.59\textwidth]{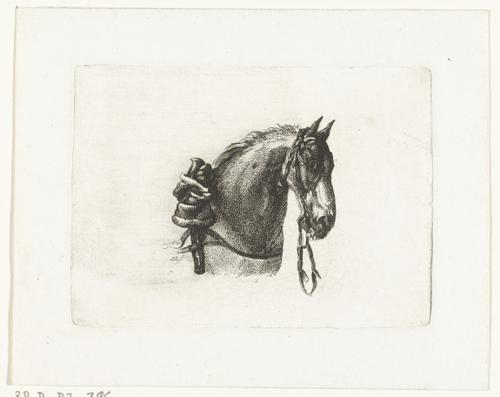}
    \caption{Recommendations for ``The Hunting With Falcon'' based on detected codes (horse, human; classification missed falcon and dog). Both \emph{hierarchy-based similarity} and \emph{IDF-weighted overlap} returned \emph{IIHIM\_-467547872.jpg} file (left). \emph{Jaccard similarity} returned \emph{IIHIM\_RIJKS\_-2107924074.jpg} (right).}
    \label{fig:hunt_recs}
\end{figure}

\begin{figure}[t]
    \centering
    \includegraphics[width=.49\textwidth]{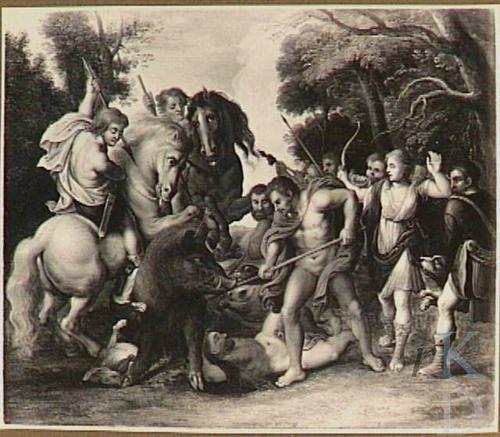}
    \includegraphics[width=.49\textwidth]{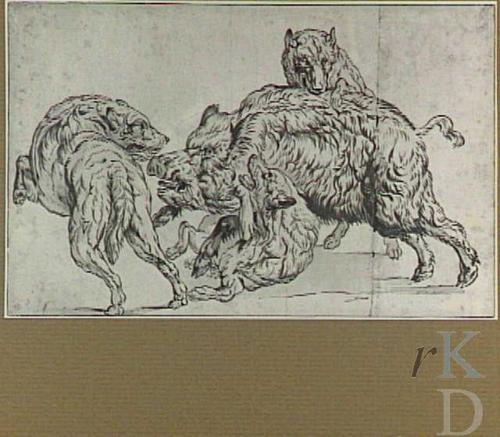}
    \caption{Recommendations for ``The Hunting With Falcon'' based on manual codes: dog, horse (+ man and animal), hoofed animals. Both \emph{hierarchy-based similarity} and \emph{IDF-weighted overlap} returned \emph{IIHIM\_1579845581.jpg} file (left). \emph{Jaccard similarity} returned \emph{IIHIM\_891269882.jpg} (right).}
    \label{fig:hunt_recs_manual}
\end{figure}

\begin{figure}[t]
    \centering
    \includegraphics[width=.5\textwidth]{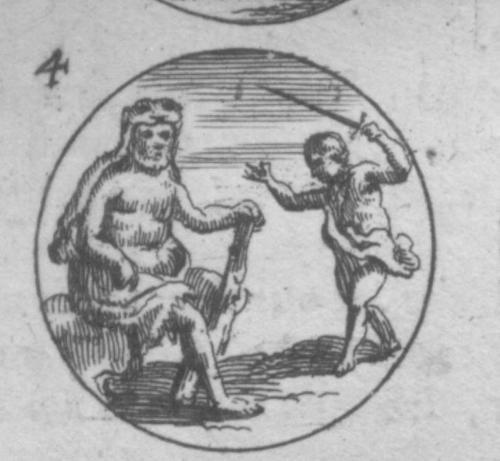}
    \caption{Recommendations for ``94L53 Hercules discovers Tiryns' famous dye: the muzzle of Hercules' dog is stained with purple after it has bitten into a mollusc'' code. Only \emph{hierarchy-based similarity} method was able to recommend an image from the Iconclass AI Test Set: \emph{embepu\_f1691\_pic0304.jpg}.}
    \label{fig:hercules}
\end{figure}

First, an evaluation was performed on the classification results for the two examples described in Sect.~\ref{sec:evaluation:classification}.
The portrait of a dog (see Fig.~\ref{fig:dog}) with the assigned code ``34B11 dog'' results in recommendations that favor canine portraits or scenes in which ``34B11'' is prominent, which is consistent with the assumptions.
The classification of the hunt scene (see Fig.~\ref{fig:hunt}) missed falcon and dog codes. Therefore, recommenders returned horse-centric scenes (see Fig.~\ref{fig:hunt_recs}). 
However, if the classification was successful and the system returned, e.g., the following set of codes: $\{$``34B11 dog'', ``46C13141(+78)  horse (+ man and animal)'', ``25F24 hoofed animals''$\}$, the narration would be better reflected in the codes, which would improve the recommendations, as presented in Fig.~\ref{fig:hunt_recs_manual}.

The third example of a recommendation is an attempt to find images matching the previously mentioned code ``94L53 Hercules discovers Tiryns' famous dye: the muzzle of Hercules' dog is stained with purple after it has bitten into a mollusc''. This is a rather unusual term referring to a specific myth. After running the recommendation module, it turns out that only \emph{hierarchy-based similarity} method was able to return an image, which depicts Hercules (see Fig.~\ref{fig:hercules}).
This is because no image in the collection has the code ``94L53'' assigned to it.
The taxonomy-based method matched the artwork with Hercules since it had the hierarchically close codes: ``94L5 non-aggressive, friendly or neutral activities and relationships of Hercules'', ``94L8(CLUB) attributes of Hercules: club'' and ``94L8(LION'S SKIN) attributes of Hercules: lion's skin''.
The presence of these codes explains the correctness of the recommendation and confirms its compliance with the assumptions.

Further evaluation was carried out by manually calling the recommendation function for different sets of codes, both corresponding to graphics from the test set and combinations of codes that are not assigned to any of the works.
These spot checks reveal complementary strengths:
\begin{enumerate}
    \item The hierarchical method succeeds when exact code matches are absent, but nearby branch codes signal the same narrative.
    \item IDF excels for queries containing rare codes. It can outweigh many common object overlaps.
    \item Jaccard is robust against candidates with many generic codes and prefers concentrated overlaps.
\end{enumerate}

\section{Summary and future work}
\label{sec:summary}

We proposed a proof-of-concept CARIS system implementing a four-stage workflow for classification and recommendation based on the Iconclass system.
The evaluation demonstrates the potential of this solution: Iconclass-aware computer vision and recommendation methods can \emph{accelerate} cataloging and \emph{enhance} navigation in large heritage repositories.
The key insight is to let computer vision propose \emph{visible} elements and to use symbolic structures (Iconclass hierarchy) to reach \emph{meaning}.

As indicated in the introduction, sustainable AI for cultural heritage requires shared vocabularies and transparent rules.
Basing the proposed pipeline on the Iconclass system, rather than solely on visual clues, addresses this need.
The Iconclass also provides a semantic backbone.
Its taxonomy establishes a structured hierarchy that encodes proximity.
Thus, it is possible to recommend not only objects with the same meaning (exactly the same codes), but also those with a similar one.

To the best of the authors' knowledge, this is the first recommendation system based on Iconclass tags described in the literature and one of the first solutions using YOLO for iconography classification.
Although the evaluation showed the potential of the technology used, more engineering is needed to make it useful to collection curators and their audiences.

The biggest limitation is the observation that end-to-end results are very sensitive to object detection quality.
Therefore, it will be crucial to create Iconclass-compliant training sets based on benchmarks annotated by experts to fine-tune YOLO models.
Future systems can also combine multimodal information---textual metadata, iconographic labels, neural image features, and even vision-language embeddings like CLIP---to provide better matched code sets for images.

The second area requiring further work is the refinement of the rule engine to facilitate the inference of more abstract Iconclass codes.
Rules can be mined semi-automatically from large Iconclass-labeled corpora (with frequent pattern mining over code co-occurrences), and then validated with curators/experts.

Finally, an end-user interface with an explainability layer is needed in order to make the system accessible to a wider audience.
Although hierarchy-aware similarity aids in interpretation, more work is needed to explain \emph{why} a code or recommendation was chosen in curator- and user-friendly terms.
This could also lead to the creation of a unified recommender that combines hierarchical, IDF, and Jaccard into a meta-ranker that selects the best strategy per query based on quality signals (e.g., rarity profile, code depth distribution).

Overall, the presented results on the classification and recommendation of artworks are in line with modern research trends pointing to a growing convergence between deep learning, multimodal semantic modeling, and user experience design, which increases the ability of digital systems to facilitate meaningful engagement with cultural heritage.

\begin{credits}
\subsubsection{\ackname}
We would like to express our sincere gratitude to Marek Pędziwiatr from the Centre for Brain Research at the Jagiellonian University, whose seminar on human vision and the processing of complex scenes as composed of individual objects inspired us to undertake the reported research.

This publication was funded by a flagship project ``CHExRISH: Cultural Heritage Exploration and Retrieval with Intelligent Systems at Jagiellonian University'' under the Strategic Programme Excellence Initiative at Jagiellonian University.
The research for this publication has been supported by a grant from the Priority Research Area DigiWorld under the Strategic Programme Excellence Initiative at Jagiellonian University.

During the preparation of this work the authors used MS Copilot and Writefully in order to improve the readability and language of the manuscript. After using these tools, the authors reviewed and edited the content as needed and take full responsibility for the content of the published article.

\subsubsection{\discintname}
The authors have no competing interests to declare that are
relevant to the content of this article.
\end{credits}

\bibliographystyle{splncs04}
\bibliography{geistbib/culheripub,geistbib/culheriteam}

\end{document}